\documentclass[12pt,preprint]{aastex}

\shorttitle{Alfv\'en waves and coronal rain}
\shortauthors{P. Antolin \& K. Shibata}

\begin{document}

\title{Coronal rain as a marker for coronal heating mechanisms}

\author{P. Antolin\altaffilmark{1,2}, K. Shibata\altaffilmark{1}}
\affil{\altaffilmark{1}Kwasan Observatory, Kyoto University,
    Yamashina, Kyoto, 607-8471, Japan}
\affil{\altaffilmark{2}The Institute of Theoretical Astrophysics, University of Oslo, P.O. Box 1029, Blindern, NO-0315 Oslo, Norway}
\email{antolin@kwasan.kyoto-u.ac.jp, shibata@kwasan.kyoto-u.ac.jp}
\altaffiltext{1}{Also at: Center of Mathematics for Applications, University of Oslo, P.O. Box 1053, Blindern, NO-0316, Oslo, Norway}

\begin{abstract}

Reported observations in H$\alpha$, Ca II H and K or or other chromospheric lines of coronal rain trace back to the days of the Skylab mission. Offering a high contrast in intensity with respect to the background (either bright in emission if observed at the limb, or dark in absorption if observed on disk) these cool blobs are often observed falling down from high coronal heights above active regions. A physical explanation for this spectacular phenomenon has been put forward thanks to numerical simulations of loops with footpoint concentrated heating, a heating scenario in which cool condensations naturally form in the corona. This effect has been termed ``catastrophic cooling'' and is the predominant explanation for coronal rain. In this work we further investigate the link between this phenomenon and the heating mechanisms acting in the corona. We start by analyzing observations of coronal rain at the limb in the Ca II H line performed by the \textit{Hinode} satellite. We then compare the observations with 1.5-dimensional MHD simulations of loops being heated by small-scale discrete events concentrated towards the footpoints (that could come, for instance, from magnetic reconnection events), and by Alfv\'en waves generated at the photosphere. It is found that if a loop is heated predominantly from Alfv\'en waves coronal rain is inhibited due to the characteristic uniform heating they produce. Hence coronal rain may not only point to the spatial distribution of the heating in coronal loops but also to the agent of the heating itself. We thus propose coronal rain as a marker for coronal heating mechanisms. 

\end{abstract}

\keywords{Sun: corona -- Sun: flares -- MHD -- waves}

\section{Introduction}

Coronal loops are dynamical entities which exhibit heating and cooling processes constantly. Most of them are actually far from being in hydrostatic equilibrium \citep{Aschwanden_2001ApJ...550.1036A}. The dynamical nature of loops can manifest in observations in a variety of forms, among which propagating intensity variations is a common one. Indeed, intensity variations traveling along coronal loops have been frequently observed in many different wavelengths, in hot coronal EUV lines as well as chromospheric cool lines. The agents producing these intensity variations can be either propagating waves or flows along coronal loops, and observers can have a hard time differentiating them despite their very different physical nature. Coronal rain is an example of intensity variations caused by flows. This spectacular phenomenon corresponds to cool plasma condensations (with chromospheric temperatures) in a hot environment (coronal temperatures) falling down along coronal loops, hence the very appropriate name. Coronal rain seems to be a common phenomenon above active regions, where loops are dense. Due to the low temperatures, the condensations appear as bright emission profiles in chromospheric lines such as H$\alpha$ or Ca II H and K when observed at the limb \citep[][Fig. \ref{fig1}]{DeGroof_2004AA...415.1141D} or as dark absorption profiles when observed on disk. If observed in EUV wavelengths these propagating blobs appear as dark features \citep{Schrijver_2001SoPh..198..325S}.

\citet{DeGroof_2004AA...415.1141D, DeGroof05} report simultaneous observations in EIT 304 \AA, H$\alpha$ images from Big Bear Solar Observatory, and in 171 \AA~ (\textit{TRACE}) of intensity variations propagating along a coronal loop. Following a detailed analysis they put forward a series of points through which a differentiation between waves and flows (cool plasma condensations, coronal rain) can be done. Propagating waves frequently reported in EIT/\textit{SOHO} 195 \AA~ and \textit{TRACE} 171 \AA~ appear as low min-to-max intensity variations as compared to coronal rain. Most observed waves correspond to sound modes propagating at constant speed corresponding to the local sound speed of the corona, $\sim$ 150 km s$^{-1}$. On the other hand, cool plasma condensations are seen to accelerate while falling along coronal loops to velocities above 100 km s$^{-1}$. Furthermore, waves have only been seen to propagate upwards, within the first 20 Mm of loops, after which they are usually damped, while cool condensations have only been seen to fall down along the loops from coronal heights. The falling speeds of the condensations have been reported to be lower than free fall speeds, due probably to the gas pressure along the loop, which increases considerably below the transition region. In some cases continuous flows from one footpoint to the other are also observed. 

\citet{Antiochos_1999ApJ...512..985A} propose a common physical mechanism for coronal rain and prominence formation. Ranges for temperatures and densities seem to be shared by both mechanisms, as well as the location of formation, high in the corona. While a prominence forms and is supported by the magnetic field for a time scale of days, coronal rain occurs on a time scale of minutes. It was shown that prominences may not need to rely on the geometry of the magnetic field to form (presence of ``dips''), contrary to the general belief. They showed that a coronal loop whose heating is concentrated towards the footpoints is subject to a thermal instability in the corona, which dramatically cools down to chromospheric temperatures in time scales of minutes once the density and temperatures have reached critical values. This phenomenon was termed ``catastrophic cooling'' and has so far gained acceptance as possible explanation for coronal rain. This mechanism occurs in a cyclic manner, matching reported periodicity of coronal rain observations \citep{Schrijver_2001SoPh..198..325S}. This periodicity depends on geometrical aspects such as the loop length and the heating scale height, and can be obtained even with a time independent heating \citep{Muller_2003AA...411..605M}. As the heating is concentrated towards the footpoints a larger mass flux is input into the corona as compared to uniform heating. The density of the corona increases in time while the energy flux is kept constant due to the constant heating input at the footpoints. This causes the loop to reach a maximum temperature after which it starts decreasing due to the decreasing of the heating per unit mass. The lower temperatures and higher densities increase the radiative losses which further cool the corona. Eventually, in a time scale of hours, the loop reaches a critical state and a thermal instability sets in. Then, in a time scale of minutes the temperature and density respectively decrease and increase dramatically to chromospheric values. The cool condensation subsequently falls down subject to gravity and plasma pressure. The loop subsequently is depleted and reheats rapidly due to the low density. The cycle then repeats. We will refer to these cycles as ``limit cycles", as termed by \citep{Muller_2003AA...411..605M}. 

Observations seem to indicate that Coronal rain is a fairly common phenomenon of active regions, where loops are hot and dense. The coronal heating mechanism being a great unknown in solar physics, it is interesting to think about its underlying relationship to coronal rain. At a first glance coronal rain may appear as a random failure of the coronal heating mechanism to heat the loops in active regions. However, as previously stated, this phenomenon seems to act in the corona in a cyclic manner. Furthermore simulations have pointed out to the necessity of specific spatially dependent heating in order to allow the catastrophic cooling leading to coronal rain \citep{Antiochos_1999ApJ...512..985A, Muller_2003AA...411..605M}. This cooling phenomenon seems then to be deeply linked to the unknown coronal heating mechanism. Since it is a fairly easily observable phenomenon due to the large velocities and density variation (hence clear Doppler shifted emission or absorption profiles), it may act as a marker of the operating heating mechanism in the loop. It has been shown for instance that footpoint concentrated heating may lead to catastrophic cooling if the heating scale height is sufficiently concentrated towards the footpoints \citep{Muller_2003AA...411..605M}. Uniform heating on the other hand fails to reproduce the phenomenon since the heating rate per unit mass needs to decrease in time locally in the corona in order to allow the thermal instability to set in. It was further shown that catastrophic cooling does not need time dependent heating. In other words, it can happen even with a constant heating function. The footpoint concentrated heating function to which simulations point to matches the observational evidence of coronal loops above active regions for being mainly heated at their footpoints \citep{Aschwanden_2001ApJ...560.1035A}, which sets most loops out of hydrostatic equilibrium. Further evidence of this fact has been found by \citet{Hara_2008ApJ...678L..67H} using the Hinode/EIS instrument, which shows that active region loops exhibit upflow motions and enhanced nonthermal velocities. Possible unresolved high-speed upflows were also found, fitting in the footpoint concentrated heating scenario. 

Many heating mechanisms have been proposed as candidates for heating the solar corona up to the observed few million degree temperatures. In this context a large emphasis has been put on the search for Alfv\'en waves in the solar corona. Theoretically, they can be easily generated in the photosphere by the constant turbulent convective motions, which inputs large amounts of energy into the waves \citep{Muller_1994AA...283..232M, Choudhuri_1993SoPh..143...49C}. Having magnetic tension as its restoring force the Alfv\'en waves can travel less affected by the large transition region gradients with respect to other modes. Also, when traveling along thin magnetic flux tubes they are cut-off free since they are not coupled to gravity (Musielak et al. 2007)\footnote{\citet{Verth_etal_aap_09} have pointed out however that the assertion made by \citet{Musielak_2007ApJ...659..650M} is valid only when the temperature in the flux tube does not differ from that of the external plasma. When this is not the case a cut-off frequency is introduced.}. Alfv\'en waves generated in the photosphere are thus able to carry sufficient energy into the corona to compensate the losses due to radiation and conduction, and, if given a suitable dissipation mechanism, heat the plasma to the high million degree coronal temperatures \citep{Uchida_1974SoPh...35..451U, Wentzel_1974SoPh...39..129W, Hollweg_1982SoPh...75...35H, Poedts_1989SoPh..123...83P, Ruderman_1997AA...320..305R, Kudoh_1999ApJ...514..493K, Antolin_2009arXiv0910.0962v1} and power the solar wind \citep{Suzuki_2006JGRA..11106101S, Cranmer_2007ApJS..171..520C}. The main problem faced by Alfv\'en wave heating is actually to find a suitable dissipation mechanism. Being an incompressible wave it must rely on a mechanism by which to convert the magnetic energy into heat. Several dissipation mechanisms have been proposed, such as parametric decay \citep{Goldstein_1978ApJ...219..700G, Terasawa_1986JGR....91.4171T}, mode conversion \citep{Hollweg_1982SoPh...75...35H, Kudoh_1999ApJ...514..493K, Moriyasu_2004ApJ...601L.107M}, phase mixing \citep{Heyvaerts_1983AA...117..220H, Ofman_2002ApJ...576L.153O}, or resonant absorption \citep{Ionson_1978ApJ...226..650I, Hollweg_1984ApJ...277..392H, Poedts_1989SoPh..123...83P, Erdelyi_1995AA...294..575E}. The main difficulty lies in dissipating sufficient amounts of energy in the correct time and space scales. For more discussion regarding this issue the reader can consult for instance \citet{Klimchuk_2006SoPh..234...41K}, \citet{Erdelyi_2007AN....328..726E} and \citet{Aschwanden_2004psci.book.....A}. Works considering Alfv\'en wave heating as coronal heating mechanism have shown that the obtained coronae are uniformly heated \citep{Moriyasu_2004ApJ...601L.107M, Antolin_2008ApJ...688..669A, Antolin_2009arXiv0910.0962v1, Suzuki_2006JGRA..11106101S}. In these works the heating issues from shocks of longitudinal modes (mainly slow modes) from mode conversion of the Alfv\'en waves due to the density fluctuation, wave-to-wave interaction and deformation of the wave shape during propagation. The coronal loops issuing from Alfv\'en wave heating are found to satisfy quite well the RTV scaling law \citep{Rosner_1978ApJ...220..643R} which quantifies the heating uniformity in the loops. This result would point then towards an inhibition of coronal rain if Alfv\'en wave heating is predominant in the loop. It is this idea that is addressed in this work. 

Another promising coronal heating candidate mechanism is the nanoflare reconnection heating model. The nanoflare reconnection process was first suggested by \citet{Parker_1988ApJ...330..474P}, who considered a magnetic flux tube as being composed by a myriad of magnetic field lines braided into each other by continuous footpoint shuffling. Many current sheets in the magnetic flux tube would be created randomly along the tube that would lead to many magnetic reconnection events, releasing energy impulsively and sporadically in small quantities of the order of $10^{24}$ erg or less (nanoflares). Parker's original idea was a nanoflare reconnection heating acting uniformly in the corona but it was later proposed to be concentrated towards the footpoints of loops, where the magnetic canopy lies and magnetic field lines may entangle \citep{Klimchuk_2006SoPh..234...41K, Aschwanden_2001ApJ...560.1035A}. In the reconnection scenario waves are also expected to be generated, and the energy imparted into Alfv\'en waves is a matter of debate. The imparted energy may well depend on the location in the atmosphere of the reconnection event. \citet{Parker_1991ApJ...372..719P} suggested a model in which 20~\% of the energy released by reconnection events in the solar corona is transfered as a form of Alfv\'en wave.\citet{Yokoyama_1998ESASP.421..215Y} studied the problem simulating reconnection in the corona, and found that less than 10~\% of the total released energy goes into Alfv\'en waves. This result is similar to the 2-D simulation results of photospheric reconnection by \citet{Takeuchi_2001ApJ...546L..73T}, in which it is shown that the energy flux carried by the slow magnetoacoustic waves is one order of magnitude higher that the energy flux carried by Alfv\'en waves. On the other hand, recent simulations by Kigure et al. (private communication) show that the fraction of Alfv\'en wave energy flux in the total released magnetic energy during reconnection in low $\beta$ plasmas may be significant (more than 50~\%). Since the observed ubiquitous intensity bursts (nanoflares) are thought to play an important role in the heating of the corona \citep{Hudson_1991SoPh..133..357H} and since they are generally assumed to be a signature of magnetic reconnection it is then crucial to determine the energy going into the Alfv\'en waves during the reconnection process. Moreover, \citet{Moriyasu_2004ApJ...601L.107M} has shown that the observed spiky intensity profiles due to impulsive releases of energy may actually be a signature of Alfv\'en waves. It was found that due to nonlinear effects Alfv\'en waves can convert into slow and fast magnetoacoustic modes which then steepen into shocks and heat the plasma to coronal temperatures balancing losses due to thermal conduction and radiation. The shock heating due to the conversion of Alfv\'en waves was found to be episodic, impulsive and uniformly distributed throughout the corona, producing an X-ray intensity profile that matches observations. Hence, \citet{Moriyasu_2004ApJ...601L.107M} proposed that the observed nanoflares may not be directly related to reconnection but rather to Alfv\'en waves. 

Differentiating Alfv\'en wave heating from nanoflare reconnection heating during observations is one of the main tasks needed in order to solve the coronal heating problem. Following the work of \citet{Moriyasu_2004ApJ...601L.107M}, \citet{Antolin_2008ApJ...688..669A} have compared both heating mechanisms by studying the hydrodynamic response of a loop subject to both kinds of heating. It was found that Alfv\'en waves lead to a dynamic, uniformly heated corona with steep power law indexes (issuing from statistics of heating events) while nanoflare-reconnection heating leads to lower dynamics (besides the times when catastrophic cooling takes place in the case of footpoint concentrated heating) and shallow power laws. It was further found that footpoint nanoflare heating (namely, nanoflare reconnection heating concentrated towards the footpoints of loops) leads to hot upflows (as observed in the Fe\,XV 284.16~\AA\ line) due to the plasma being heated rapidly towards the footpoints before being ejected into the corona, while Alfv\'en wave heating leads to hot downflows due to the plasma achieving the maximum temperatures in the corona (rather than at the footpoints) and being carried back to the footpoints by the strong shocks generated there \citet{Antolin_2009arXiv0903.1766A}. In this work we propose coronal rain as another observational signature through which both heating mechanisms can be distinguished. 

We start first by reporting limb observations of coronal rain from \textit{Hinode}/SOT in the Ca\,II H line. The velocities and shapes of the falling condensations are analyzed. With a 1.5-dimensional code we then proceed to model a coronal loop being subject to a heating mechanism that is concentrated towards the footpoints, such as the nanoflare reconnection heating model \citep[as proposed by][]{Aschwanden_2001ApJ...560.1035A, Klimchuk_2006SoPh..234...41K}. In the case considered catastrophic cooling happens 3 times and we select the first event to analyze and compare with our observations. Next we proceed by generating Alfv\'en waves at the photosphere and gradually increase the amplitude. As the heating from the waves becomes non-negligible with respect to the heating flux from the nanoflare heating events catastrophic cooling is inhibited and the loop reaches a thermal equilibrium state. This result has important consequences since it points to the conclusion that Alfv\'en waves are a non predominant heating mechanism in active region loops.

The work is organized as follows. In \S\ref{observations} we report observations of coronal rain in the Ca II H line performed with \textit{Hinode}/SOT. In \S\ref{model} we introduce the 1.5-dimensional MHD model in which our loop is based on and discuss the heating models of the loop. In \S\ref{simulation} we present the results of footpoint heating and analyze a typical case of catastrophic cooling. The effect of Alfv\'en waves on the thermal stability of the loop is also studied. In \S\ref{discussion} we discuss the results in the context of coronal heating and conclude the work.

\section{Observations of coronal rain with Hinode/SOT}\label{observations}

\begin{figure}
\epsscale{1.}
\plotone{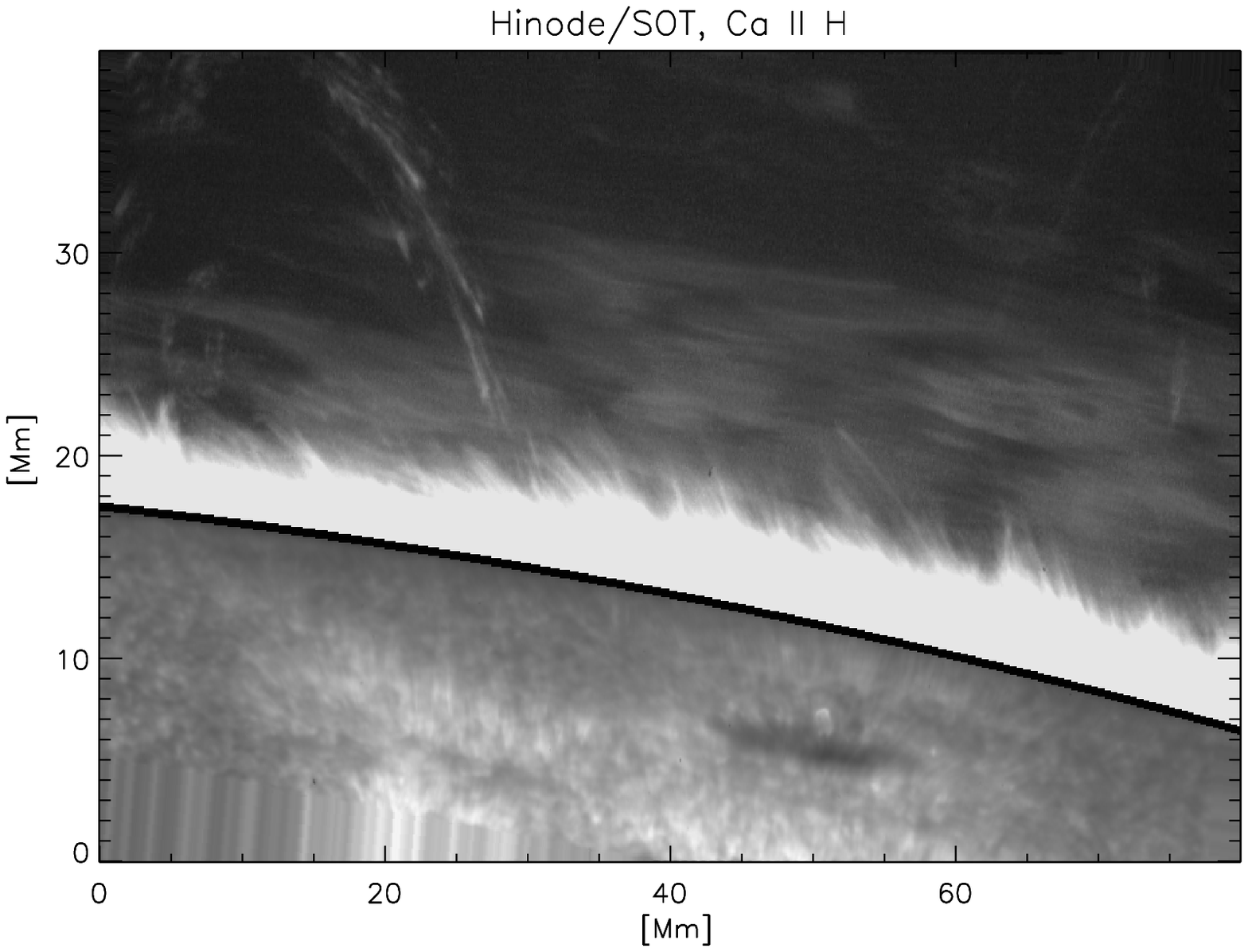}
\caption{High resolution Hinode/SOT image at the limb in the Ca II H line. The entire field of view corresponds to roughly 80 Mm $\times$ 40 Mm. The observation was performed on November 9, 2006, between 19:33 to 20:44 UT with a cadence of 15 s. In order to be able to see the fainter corona the intensity of the photosphere (disk) has been decreased. Also, the intensity of the limb structures (spicules) is saturated. The sunspot visible on disk corresponds to the main sunspot of NOAA AR 10921. Above the limb, cloud-like prominence structures and coronal rain can be observed. The intensity of coronal rain in the Ca II H line is about 1$\%$ of the on-disk photospheric intensity. \label{fig1}}
\end{figure}

In this section we report high resolution observations of coronal rain at the limb performed by the Solar Optical Telescope on board of Hinode in a 0.3-nm broadband region centered at 396.8 nm, the H-line spectral feature of singly ionized calcium (Ca II H line). The observation was performed on 9 November 2006 from 19:33 to 20:44 UT with a cadence of 15 s, and focused on NOAA AR 10921 on the west solar limb. The corresponding field of view is 80 Mm $\times$ 40 Mm. The Ca II H line is a chromospheric line typically showing plasmas with temperatures on the order of 20,000 K. In the data set, apart from the active region on disk, many interesting structures can be seen over the limb, such as spicules, prominences and coronal rain (Fig.~\ref{fig1}). The same data set was used by \citet{Okamoto_2007Sci...318.1577O} to report observations of transverse magnetohydrodynamic waves propagating in the observed prominences. The later show complex horizontal threads displaying continuous horizontal motions, and appear to be located on the background of the images. Coronal loops exhibiting coronal rain appear to be located on the foreground of the prominences. The cool condensations falling down the loops form close to the apex at heights of $\sim40$ Mm above the surface, and do not seem to be linked to the prominences.

\begin{figure}
\epsscale{1.}
\plotone{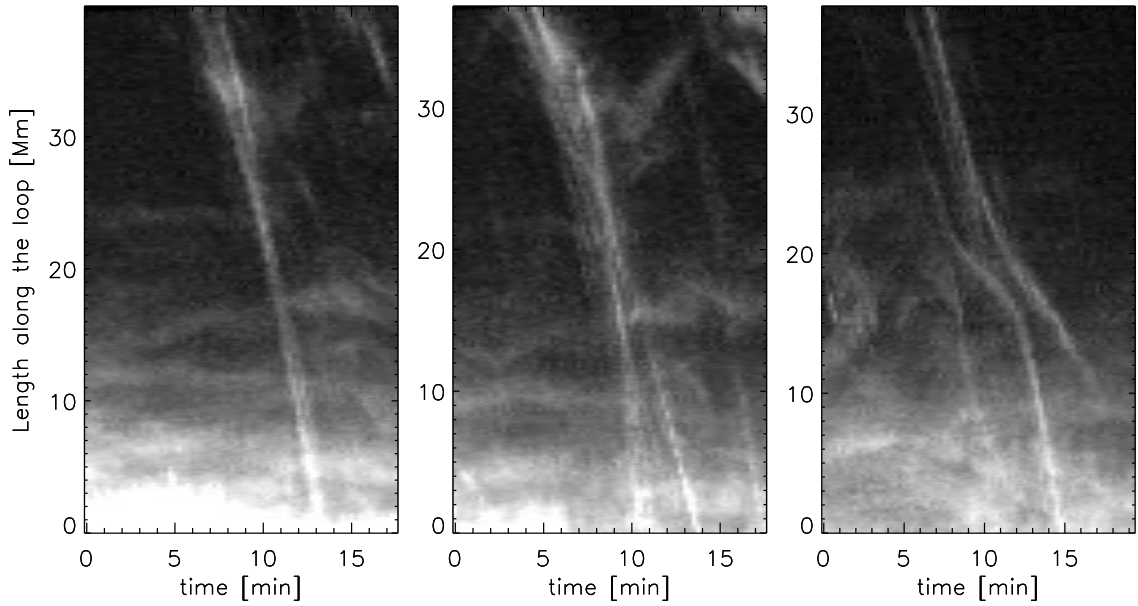}
\caption{Length-time diagrams for 3 loops exhibiting coronal rain on the upper left of Fig. \ref{fig1}. The loops are traced by quadratic interpolation. The $y$ axis denotes the length along the loops. Coronal rain, which appears brighter in the Ca II H line intensity, generally accelerates downward under the action of gravity. However, deceleration and even upward motions can also be observed (right and middle panels respectively).    \label{fig2}}
\end{figure}

The condensations conforming coronal rain trace the geometry of the coronal loops and appear to accelerate as they fall down to the chromosphere under the action of gravity. By fitting the loops with a curve we can trace the condensations along their way down and calculate the corresponding velocities. In Fig.~\ref{fig2} we show length-time diagrams of 3 loops exhibiting coronal rain located on the upper left side of Fig.~\ref{fig1}. The estimated velocities on the top part of the loops (near the apex), and close to the footpoints are displayed in Table~\ref{table3.1}. We also calculate the free fall velocity that the condensations would have under the action of gravity alone and no initial velocity (at the top part of the loops). We can see that close to the top parts of the loop the condensations have velocities around 50 - 70 km s$^{-1}$, from which we estimate the apexes of the loops to be at 5 - 10 Mm above the top of the figures (assuming that the condensations start with zero velocity at the apex). The lengths of these loops should then be close to 100 Mm. At the footpoints of the loops the condensations have their highest velocities, roughly between 115 and 140 km s$^{-1}$. The calculated free fall velocities seem to match these values, however, they have been calculated assuming no initial velocity. We can then expect a discrepancy on the order of the velocity at the top ($\sim50$ km s$^{-1}$). Hence, the condensations fall at velocities which are lower than the free fall velocities, as has been reported previously for coronal rain observations \citep{DeGroof_2004AA...415.1141D}. The decelerations may be caused by internal pressure changes in the loops, as suggested by the results of the simulations reported in section \ref{simulation}. For the loop in the right panel we have a clear deceleration of the coronal rain from $\sim70$ to $\sim15$ km s$^{-1}$ halfway down along the loop. It then seems to accelerate again to a speed above 100 km s$^{-1}$, higher than the corresponding free fall velocity. For the loop in the middle panel we can see an upward motion of coronal rain at $\sim25$ km s$^{-1}$ close to the top part of the loop. This may be caused by a cooling and heating process acting upward in that part of the loop, thus resulting in an apparent upward motion of the coronal rain. However, it may also be a true motion caused by a pressure change inside the loop at that height, such as from propagating acoustic shocks, as the simulations performed in \ref{simulation} suggest.

\section{Simulation setup}\label{model}

In order to simulate coronal rain we follow the model of \citet{Antiochos_1999ApJ...512..985A}, in which it is shown that cool condensations can dynamically form in the corona resulting from footpoint heating of the loops. This mechanism is known as catastrophic cooling and is further explained in this section. 

\subsection{Model}

We consider a magnetic flux tube (loop) of 100 Mm in length, roughly the same length as the loops exhibiting coronal rain in the observations reported here. The geometry of the loop takes into account the area, which considers the predicted expansion of magnetic flux in the photosphere and chromosphere, displaying an area ratio between the corona and the photosphere of 1000. As discussed in the introduction catastrophic cooling as proposed by \citet{Antiochos_1999ApJ...512..985A} needs the heating to be concentrated towards the footpoints of loops. In the case of footpoint heating without the generation of Alfv\'en waves in the photosphere the plasma motion is governed by the usual 1D HD equations for the conservation of mass, momentum and energy. The model in this case is the same as the heating model considered for nanoflare reconnection with heating concentrated towards the footpoints in \citet{Antolin_2008ApJ...688..669A}, and is further discussed below. When Alfv\'en waves are considered the model gains the azimuthal component and is the same model as the model for Alfv\'en wave heating in \citet{Antolin_2008ApJ...688..669A}.

We write the 1D HD equations for the conservation of mass, momentum and energy in the following way:

\noindent the mass conservation equation:
\begin{equation}\label{mass}
\frac{\partial\rho}{\partial t}+v\frac{\partial\rho}{\partial s}=
    -\rho B\frac{\partial}{\partial s}\left(\frac{v}{B}\right);
\end{equation}
the momentum equation:
\begin{equation}\label{momentums}
\frac{\partial v}{\partial t}+v\frac{\partial v}{\partial s}=
    -\frac{1}{\rho}\frac{\partial p}{\partial s}-g_{s};
\end{equation}
and the energy equation:
\begin{equation}\label{energy}
\frac{\partial e}{\partial t}+v\frac{\partial e}{\partial s}=
    -(\gamma-1)e B\frac{\partial}{\partial s}\left(\frac{v}{B}\right)
    -\frac{R-\mathcal{S}-\mathcal{H}}{\rho}+\frac{1}{\rho r^{2}}\frac{\partial}{\partial
    s}\left(r^{2}\kappa\frac{\partial T}{\partial s}\right);
\end{equation}
where 
\begin{equation}\label{glawinte}
        p=\rho\frac{k_{B}}{m}T, \hspace{0.5cm} e=\frac{1}{\gamma-1}\frac{p}{\rho}.
\end{equation}
In the above equations (\ref{mass})-(\ref{energy}) $s$ measures the distance along the flux tube (central field line) and $r$ is the radius of the tube. $\rho$, $p$, $v$ and $e$ are, respectively, density, pressure, velocity along the loop and internal energy; $B$ is the magnetic field along the loop and is a function of $r$ alone, $B=B_{0}(r_{0}/r)^{2}$, where $B_{0}$ is the value of the magnetic field at the photosphere and $r_{0}=200$ km is the initial radius of the loop; $k_{B}$ is the Boltzmann constant and $\gamma$ is the ratio of specific heats for a monatomic gas, taken to be 5/3. The $g_{s}$ is the effective gravity along the loop and is given by
\begin{equation}\label{gravitys}
    g_{s}=g_{\odot}\cos\left(\frac{s}{L}\pi\right),
\end{equation}
where $g_{\odot}=2.74\times10^{4}$ cm s$^{-2}$ is the gravity at the base and $L$ is the total length of the loop.

We assume an inviscid perfectly conducting fully ionized plasma. The effects of thermal conduction and radiation are taken into account, where the Spitzer conductivity corresponding to a fully ionized plasma is considered, and radiative losses are defined as 
\begin{equation}\label{opticalthin}
    R(T)=n_{e}n_{p}Q(T)=\frac{n^{2}}{4}Q(T).
\end{equation}
Here, $n=n_{e}+n_{p}$ is the total particle number density ($n_{e}$ and $n_{p}$ are, respectively, the electron and proton number densities, and we assume $n_{e}=n_{p}=\rho/m$ to satisfy plasma neutrality, with $m$ the proton mass) and $Q(T)$ is the radiative loss function for optically thin plasmas \citep{Landini_1990AAS...82..229L} which is approximated with analytical functions of the form $Q(T)=\chi T^{\gamma}$. We take the same approximation as in \citet[][please refer to their Table 1]{Hori_1997ApJ...489..426H}. Apart from the optically thick regions in the low solar atmosphere, it is possible to have optically thick plasma in some frequencies high up in the corona by means of catastrophic cooling of loops which would lead to coronal rain. The dynamical state of the plasma in this situation (catastrophic cooling events happen in the time scale of minutes) would produce non-equilibrium ionization effects which should be taken into account by solving the ionization rate equations in a self-consistent way (by coupling the equations to the hydrodynamic equations through the energy equation). Here we are then assuming that the radiation fields in all directions and all frequencies and the level populations do not affect the ionization level of the plasma. Since coronal rain disappears in the same time scale (minutes), we may expect the plasma that becomes optically thicker not to affect considerably the energy equation. We consider however an approximation to optically thick radiation. For temperatures below $4\times10^{4}$ K we assume that the plasma inside the loop becomes optically thick. In this case, the radiative losses $R$ can be approximated by $R(\rho)=4.9\times10^{9}\rho$ \citep{Anderson_1989ApJ...336.1089A}. In equation (\ref{energy})  the heating term $\mathcal{S}$ has a constant non-zero value which is non-negligible only when the atmosphere becomes optically thick. Its purpose is mainly for maintaining the initial temperature distribution of the loop. Here $\mathcal{H}$ denotes the heating function in the loop, which corresponds to a nanoflare heating model and is presented in the next section.

For generating Alfv\'en waves in the loop we follow the same model as in \citet{Antolin_2008ApJ...688..669A}. Random torque motions are produced in the photosphere, which generates Alfv\'en waves with a white noise spectrum in frequency. We adopt this model instead of a monochromatic wave generator since we consider that the buffeting of magnetic field lines by convective motions has a turbulent nature, thus leading to random motions. For further details about this model please refer to \citet{Antolin_2008ApJ...688..669A}.

\subsection{Nanoflare heating function}

As shown by \citet{Antiochos_1999ApJ...512..985A}, in order for catastrophic cooling to happen we have to apply a heating mechanism that is concentrated towards the footpoints of the loop. There may be many proposed heating mechanisms which can act preferentially towards the footpoints of loops. Here we will assume that the loop is subject to ``footpoint nanoflare" heating from the nanoflare reconnection model described in \citet{Antolin_2008ApJ...688..669A}. In this picture we assume that the energy imparted into Alfv\'en waves from reconnection events is low and can be neglected relative to the imparted energy to the slow modes. Hence, in this picture the corona would be heated mainly by the accumulation of numerous nanoflares coming from reconnection events and by slow magnetoacoustic shocks. Hydrodynamic modeling of nanoflare heating has already been done in the past \citep{Walsh_1997SoPh..171...81W, Cargill_2004ApJ...605..911C, Patsourakos_2005ApJ...628.1023P, Taroyan_2006AA...446..315T, Mendozabriceno_2005ApJ...624.1080M}. The nanoflare model considered here is similar to the model of \citet{Taroyan_2006AA...446..315T} with respect to the heating function $\mathcal{H}$ in equation (\ref{energy}). In the present case we assume that heating events simulating reconnection events (leading to nanoflares) occur towards the footpoints of the loop. These are input randomly as artificial perturbations in the internal energy of the gas (thus generating only slow modes). The heating rate due to the nanoflares is represented as
\begin{equation}\label{nanototheat}
    \mathcal{H}=\sum_{i=1}^{n}\mathcal{H}_{i}(t,s)
\end{equation}
where $\mathcal{H}_{i}(t,s)$, i=1,...,n are the discrete episodic heating events, and $n$ is the total number of events. 
\begin{equation}\label{heatinfunc}
    \mathcal{H}_{i}(t,s)=
\left\{
       \begin{array}{ll}
       E_{0}\sin\left(\frac{\pi(t-t_{i})}{\tau_{i}}\right)
       \exp\left(-\frac{|s-s_{i}|}{s_{h}}\right), & \hbox{$t_{i}<t<t_{i}+\tau_{i}$;} \\
       0, & \hbox{otherwise,}
       \end{array}
\right.
\end{equation}
where $E_{0}$ is the maximum volumetric heating and $s_{h}$ is the heating scale length. The offset time $t_{i}$, the maximum duration $\tau_{i}$, and the location $s_{i}$ of each event are randomly distributed in the following ranges:
\begin{equation}\label{parameters}
    t_{i}\in[0, t_{total}],\hspace{0.3cm}\tau_{i}\in[0,\tau_{max}],\hspace{0.3cm}
    s_{i}\in[s_{min}, s_{max}]\bigcup[L - s_{max}, L - s_{min}],
\end{equation}
where $t_{total}$ is the total simulation time and $s_{min}$ ($s_{max}$) define the lower (upper) boundaries of the range in the loop where heating events occur.

In order to set the values to the parameters of the heating function, equations (\ref{nanototheat})-(\ref{parameters}), an estimate of the nanoflare duration time is needed. One of the hardest parameters to estimate in magnetic reconnection theory is the thickness of the current sheet, i.e. the length across the reconnection region. If this parameter is of the order of $\sim1000$ km, the timescale of a (small) reconnection event leading to a nanoflare should oscillate between 1 and 10 s, since the order of the Alfv\'{e}n speed in the chromosphere and in the corona is, respectively, $\sim100$ and $\sim1000$ km s$^{-1}$. This value, however, is not established. 
Different values have been tried for the parameters of the heating function defined in equations (\ref{nanototheat}) - (\ref{parameters}). Since the purpose of this work is not to study the ranges in which catastrophic cooling happens we will limit ourselves in the present model to present a typical case in which it happens. For this case we have the maximum duration time of a heating event $\tau_{max}=40$ s, the heating scale length $s_{h}=1000$ km, the maximum volumetric heating $E_{0}=0.5$ erg cm$^{-3}$ s$^{-1}$ , a frequency of heating events of 1 each 50 s, the upper and lower boundaries of the ranges in which heating occurs 
$\{s_{min}=2,s_{max}=20\}$ Mm. These parameters set a mean energy per event of $1.9\times10^{26}$ erg and a mean energy flux of $2.5\times10^{7}$ erg cm$^{-2}$ s$^{-1}$.

\subsection{Initial conditions and numerical code}

The loop is assumed to follow hydrostatic pressure balance in the subphotospheric region and in the photosphere up to a height of $4H_{0}=800$ km, where $H_{0}$ is the pressure scale height at $z=0$. For the rest of the loop, density decreases as $\rho\propto h^{-4}$, where $h$ is the height from the base of the loop. This is based on the work by \citet{Shibata_1989ApJ...338..471S, Shibata_1989ApJ...345..584S}, in which the results of 2D MHD simulations of emerging flux by Parker instability exhibit such pressure distribution. The initial temperature all along the loop is set at $T=10^{4}$ K. The density at the photosphere ($z=0$) is set at $\rho_{0}=2.53\times10^{-7}$ g cm$^{-3}$, and, correspondingly, the photospheric pressure is $p_{0}=2.09\times10^{5}$ dyn cm$^{-2}$. The plasma $\beta$ parameter is chosen to be unity at $z=0$, setting the photospheric magnetic to $B_{0}=2.29\times10^{3}$ G. The value of the magnetic field at the top of the loop is then $B_{top}=2.29$ G.

The spatial resolution in the numerical scheme is set to 5 km up to a height of $\sim$16000 km. Then, the grid size slowly increases until it reaches a size of 20 km in the corona. The size is then kept constant up to the apex of the loop.  The total grid number is 10000. We take rigid wall boundary conditions at the photosphere.  The numerical scheme adopted is the CIP (cubic interpolated propagation) scheme \citep{Yabe_1991CoPhC..66..219Y}. Please refer to \citet{Kudoh_1998ApJ...508..186K} for details about the application of these scheme. The total time of the simulation is 568 minutes.

\section{Simulation results}\label{simulation}

We first perform the simulation of the loop being heated only by the events from the ``footpoint nanoflare'' model, that is, without Alfv\'en waves. We then allow the generation of Alfv\'en waves at the footpoints of the loop and analyze the effect on the catastrophic cooling events. 

\subsection{Footpoint nanoflare heating}

\begin{figure}
\epsscale{0.7}
\plotone{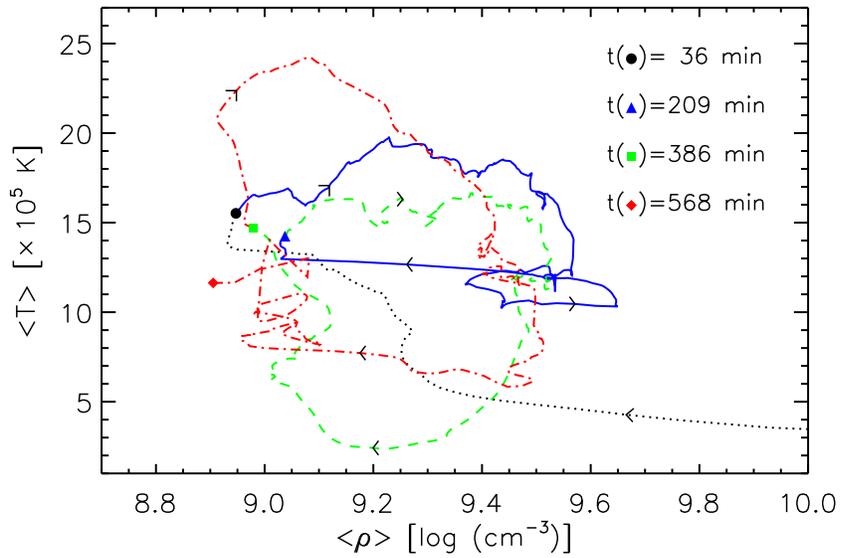}
\caption{Phase diagram of mean temperature and mean density of the corona. Arrows show the time direction and solid, dashed and dot-dashed curves denote the limit cycles (blue, green and red in the online version). The circle, triangle, square and lozenge denote the end of these cycles, respectively. The dotted curve corresponds to the start of the simulation.  \label{fig3}}
\end{figure}

Due to the large energy flux from the heating events the corona is formed rapidly in the considered footpoint nanoflare model (in about 20 min). The mean temperature in the corona over the entire simulation time is $\langle T\rangle\sim1.4\times10^{6}$ K, with a maximum temperature of $\langle T\rangle\sim3.9\times10^{6}$ K. The mean density in the corona is high, $\langle n\rangle\sim1.3\times10^{9}$ cm$^{-3}$, characteristic of a dense active region loop. As the heating events occur randomly in time we have a uniform heating input into the loop. Since the heating events occur close to the footpoints, chromospheric matter is constantly being push upwards into the loop, increasing the density in the corona. Fig.~\ref{fig3} is a phase diagram of the mean temperature and the corresponding mean density in the corona in time. Arrows indicate the time direction and curve styles indicate different cycles the loop experiences. In the present case 3 cycles can be noticed, each one lasting roughly 170 min. The dotted curve in Fig.~\ref{fig3} corresponds to the initial phase of the simulation, then the 3 cycles start, denoted by solid, dashed and dot-dashed curves (blue, green and red curves in the online version), respectively in time. A cycle is composed of 4 distinctive phases. First, a phase in which the temperature of the corona increases rapidly and the density is roughly constant. Then follows a phase of constant temperature and slow density increase. In the third phase the temperature in the corona slowly decreases and the density is roughly constant. The last phase is marked by a dramatic decrease of temperature which can happen either locally in the corona or globally (entire collapse of corona), accompanied by a dramatic increase of density (at one or more locations in the corona). These cycles have been termed ``limit cycles" by \citet{Muller_2003AA...411..605M} and can be understood as follows. 

\begin{figure}
\epsscale{1.0}
\plotone{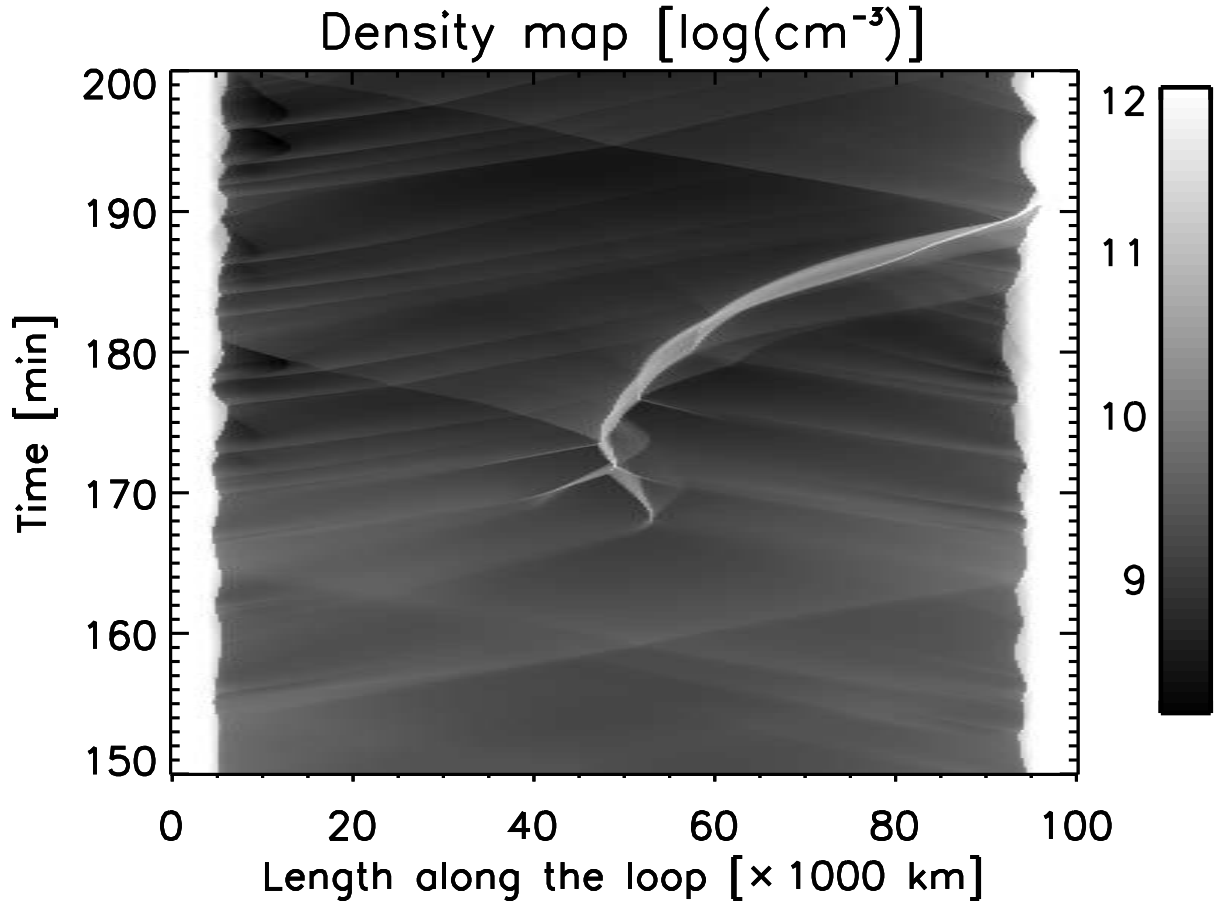}
\caption{Density map along the loop. Catastrophic cooling occurs forming a cool and dense condensation close to the apex of the loop, which falls down with increasing speed due to gravity. The motion of the condensation is also subject to the local gas pressure which varies considerably due to the acoustic shocks produced by the heating events at the footpoints of the loop. The traces of the propagating shocks are clearly observed. Note the strong deceleration of the blob as it enters the chromosphere, and the following oscillation of the transition region. \label{fig4}}
\end{figure}

Initially, the density in the corona is low, since gravity depletes the loop before having a considerable injection of mass from the heating mechanism. The corona is thus easily heated to high temperatures from the footpoint heating events (phase 1). These events continuously inject material into the corona, thus slowly increasing its density. Since the energy flux to the corona is kept constant, the high temperatures can only be kept for a specific density range (phase 2). Then the mean temperature starts decreasing due to the steady decreasing heating rate per unit mass (phase 3). The lower temperature also increases the (optically thin) radiative losses, accelerating the cooling of the corona. The loop then reaches a critical state in which its density is too high and the temperature too low. A thermal instability follows due to the large radiation increase at low temperatures (becoming optically thick) in just the same way the transition region forms (phase 4). The density (temperature) then rapidly increases (decreases) to chromospheric values in a time scale of minutes, thus leading to a catastrophic cooling event. The thermal instability can happen locally in the corona, thus forming a dense and cool blob which subsequently falls down due to gravity, or can have a more global character, case in which the entire corona collapses and several dense blobs are formed. The loop is then evacuated and gets rapidly reheated due to the low density and the constant heating input. The cycle thus restarts. In the cycle denoted by the solid line (blue line in the online version) of Fig.~\ref{fig3}  the catastrophic cooling occurs locally in the corona, while in the subsequent cycles it occurs globally. As soon as the condensation forms, its density increases rapidly and it grows larger. This is due to the constant heating events happening towards the footpoints constantly pushing plasma upwards which accumulates in the condensation. This can be seen in Fig.~\ref{fig4}, where the evolution in time of the density along the loop is shown. The (acoustic) shocks created by the heating events can be followed from the transition region. Some of them also form small condensations while propagating, before colliding with the bigger original condensation. It can be noticed that the blob experiences a change of direction at the loop top, going first towards the left footpoint and then towards the right footpoint. A shock collides with the blob and is reflected just at the time in which the blob changes direction. Another factor of the change of direction of the blob is the lower gas pressure region on the right side of the blob, as compared to the left side. Hence, these condensations are not only subject to gravity but also to the local changes of gas pressure in the loops. This mechanism may explain the observed deceleration motion and upwards motion of coronal rain in Fig.~\ref{fig2}. When the blob falls down to the chromosphere it experiences a strong deceleration by the higher density region, the transition region is left oscillating up and down a couple of times. 

\begin{figure}
\epsscale{0.7}
\plotone{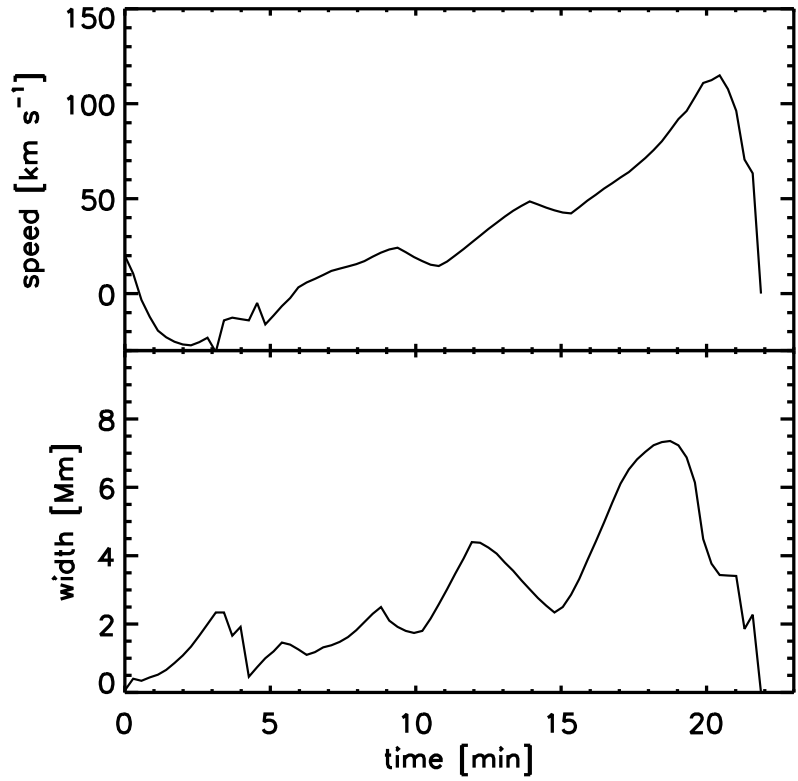}
\caption{Upper panel: velocity of the condensation along the loop with respect to time. t=0 min corresponds to the formation time of the condensation. Lower panel: Width of the condensation with respect to time since its formation time in the corona.  \label{fig5}}
\end{figure}

In the upper panel of Fig.~\ref{fig5} we track down the condensation as it falls down to the chromosphere and plot its velocity along the loop in time. The initial motion towards the left footpoint has a maximum speed of almost $\sim30$ km s$^{-1}$ before changing direction and accelerating towards the right footpoint under the action of gravity. The maximum speed is almost $120$ km s$^{-1}$ close to the footpoint, before being decelerated by the high gas pressure of the chromosphere. The obtained values for the velocity match well the velocities of coronal rain we have found from the observations with \textit{Hinode}/SOT. The lower panel shows the width of the condensation as it falls down the loop. We can see a general tendency to elongate, as some parts of the blob may accelerate faster due to the variation of the effective gravity along the loop, and also to the large accumulation of matter in the condensation. This elongation is observed for the loops exhibiting coronal rain in the observations reported here, and has also been previously reported \citep{Schrijver_2001SoPh..198..325S}. 

\subsection{Footpoint nanoflare heating and Alfv\'en waves}

\begin{figure}
\epsscale{0.7}
\plotone{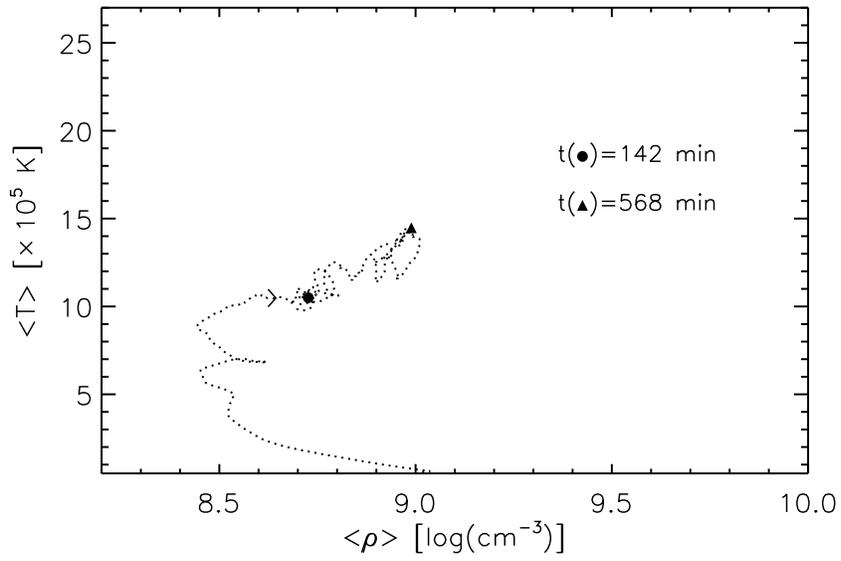}
\caption{Phase diagram of the mean temperature and density in the corona in the case of a loop heated by Alfv\'en waves. The arrow indicates the time direction. The times corresponding to the circle and the triangle (end of simulation) are indicated. Limit cycles are absent in this case. The corona reaches a uniform energy state, which acts as an attractor in the diagram.  \label{fig6}}
\end{figure}

In order to analyze correctly the effect of Alfv\'en waves on the catastrophic cooling events we first consider the case in which we only have Alfv\'en waves and no nanoflare reconnection heating. Fig.~\ref{fig6} shows the phase diagram of mean temperature and mean density of the corona in this case. We can see that as the simulation evolves the mean temperature and density in the corona converge rapidly to roughly constant values, seen as an attractor in the phase diagram. Indeed, after one fifth of the total simulation time (142 min) the temperatures and densities of the corona stay roughly constant. As seen in \citet{Antolin_2008ApJ...688..669A} the corona that issues from Alfv\'en wave heating is uniform and steady, and satisfies the RTV scaling law. 

\begin{figure}
\epsscale{0.7}
\plotone{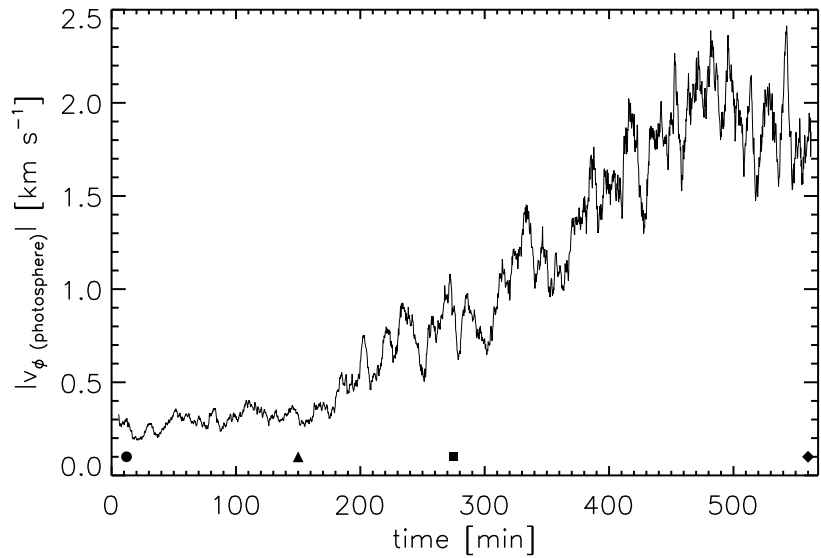}
\caption{Photospheric azimuthal (rms) velocity amplitude with respect to time. The loop is subject to both nanoflare reconnection heating and Alfv\'en wave heating. The amplitude of the Alfv\'en waves increases with time becoming a non-negligible energy source in the second half of the simulation. Symbols indicate the same times as the corresponding same symbols in Fig.~\ref{fig8}.   \label{fig7}}
\end{figure}

\begin{figure}
\epsscale{0.7}
\plotone{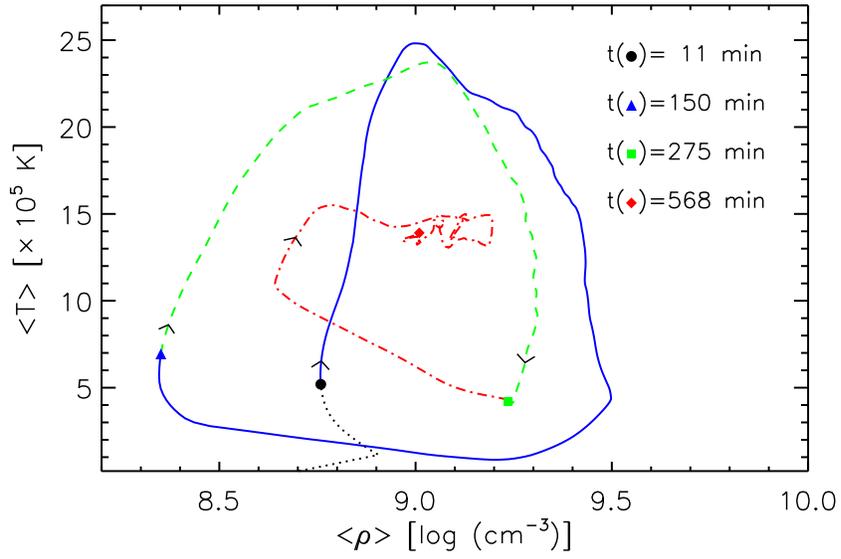}
\caption{Phase diagram of the mean temperature and density in the corona in the case of a loop with both nanoflare reconnection heating and Alfv\'en wave heating. Arrows indicate the time direction, and solid, dashed and dot-dashed curves (blue, green and red curves in the online version) denote limit cycles (the circle, triangle and square denote the start of these cycles, respectively). The dotted curve corresponds to the initial stage of the simulation. In this case the amplitude of the Alfv\'en waves increases in time as indicated in Fig.~\ref{fig7}. When the energy from the latter becomes non-negligible the corona reaches thermal equilibrium and correspondingly the cycles converge to a uniform state.  \label{fig8}}
\end{figure}

We then consider an hybrid model in which both heating mechanisms are present. Heating events simulating reconnection events happen close to the footpoints, and Alfv\'en waves with a white noise spectrum are generated in the photosphere. We further allow the amplitude of the waves to increase in time, as shown in Fig.~\ref{fig7}. At the beginning of the simulation the energy flux from the waves is negligible with respect to the energy flux from the nanoflare reconnection events. Indeed, Alfv\'en waves have an amplitude lower than 0.3 km s$^{-1}$, which from the study in \citet{Antolin_2008ApJ...688..669A} (see Fig. 4 in that paper) we know it's not enough to produce a hot corona. In the second half of the simulation the waves have an amplitude larger than 1 km s$^{-1}$ (reaching $\sim2$ km s$^{-1}$ by the end of the simulation), which, according to Fig. 4 in that paper, is enough to produce a hot corona. In Fig.~\ref{fig8} we plot the corresponding phase diagram for this case. We can see that as the amplitude of the waves increases the limit cycles get smaller and disappear. Mean temperatures and densities finally converge to roughly constant values. Hence the loop becomes uniformly heated as the heating from the Alfv\'en waves is no longer negligible.

\section{Discussion and conclusions}\label{discussion}

Observations in chromospheric lines such as H$\alpha$ or Ca II H and K seem to show that coronal rain is a phenomenon exclusively of active regions \citep{Schrijver_2001SoPh..198..325S, DeGroof_2004AA...415.1141D}, where loops are dense and heating appears to be concentrated towards the footpoints \citep{Aschwanden_2001ApJ...550.1036A, Aschwanden_2001ApJ...560.1035A, Hara_2008ApJ...678L..67H}. In this work we have reported observations with the Hinode/SOT telescope in the Ca II H line of coronal rain occurring over an active region, and compared with results of simulations of loops that undergo catastrophic cooling. We have found that catastrophic cooling of coronal loops satisfactorily reproduces the main observed features of coronal rain. The loops are preferentially heated towards the footpoints and are subject to cycles 
\citep[``limit cycles'', as termed by][]{Muller_2003AA...411..605M} in which they rapidly cool down and then reheat, they get dense and then deplete. The constant heating input at the footpoints of the loops produces coronae out of hydrostatic equilibrium. The coronal density increases in time causing a gradual decrease of the temperature and increase of the radiative losses. When the temperatures are sufficiently low radiative losses are dramatically increased since the condensation becomes optically thick and radiates considerably more. Catastrophic cooling sets in, either locally in the corona or globally, case in which the entire corona is cooled down to chromospheric values. Dense condensations of cool plasma form at coronal heights, which subsequently fall down by gravity. The loop is then depleted and the cycle starts again.

The characteristics of our condensation match well the main characteristics of the coronal rain from the reported observations. The velocities are in the same range as would be expected from coronal rain forming in roughly the same locations of loops having essentially the same length. We obtain an elongation of the condensation in the simulation as it gets denser and the effective gravity increases along its way down the loop. This elongation is observed as well in the present Hinode observations and has previously also been reported \cite{Schrijver_2001SoPh..198..325S}. The temperatures and densities of the resulting condensation have chromospheric values, which suggest that they would emit radiation in lines such as H$\alpha$ or Ca II H or K, as in the observations with Hinode/SOT or the Swedish Solar Telescope. In a future paper we will test this idea by synthesizing the H$\alpha$ line profile based on self-consistent solution of the ionization rate equations coupled with the hydrodynamic equations.

As suggested by the simulations coronal rain is subject not only to gravity but also to the pressure changes inside the loop. Catastrophic cooling is the abrupt loss of thermal equilibrium in the corona, through which large pressure changes along the loop can occur and which can drive strong flows and shocks along the loop. In the catastrophic cooling event from the simulation reported here an upward motion of the condensation is obtained due to the local changes in pressure from the flows and the shocks. This may be a possible explanation to the upward motion of coronal rain as reported from the observations with Hinode/SOT in the Ca II H line. 

If coronal rain is indeed the consequence of the catastrophic cooling mechanism we then must have a heating mechanism acting preferentially towards the footpoints in loops where coronal rain is observed, i.e. active region loops. We have seen that Alfv\'en waves generated at the footpoints of loops produce uniform coronae and are thus unable to reproduce phenomena such as coronal rain. Furthermore, when Alfv\'en waves are present in a loop and have enough energy flux to heat and maintain a hot corona the catastrophic cooling events are inhibited. The loop then converges to a uniform and steady state. Coronal loops in active regions show a recurrent occurrence of coronal rain. They are dynamical entities showing heating and cooling processes at all times. Our results indicate that having Alfv\'en wave heating as the main heating mechanism in loops means the absence of coronal rain. These results indicate then that Alfv\'en wave heating may not be the principal heating mechanism for coronal loops in active regions. 

\citet{Hara_2008ApJ...678L..67H}, using the \textit{Hinode}/EIS instrument, have found upflow motions and enhanced nonthermal velocities in the hot lines of Fe XIV 274 and Fe XV 284 in active region loops. Possible unresolved high-speed upflows were also found. In \citet{Antolin_2008ApJ...688..669A} and \citet{Antolin_2009arXiv0903.1766A} we found that footpoint or uniform heating coming from nanoflare reconnection exhibit hot upflows, thus fitting in the observational scenario of active regions, while Alfv\'en wave heating was found to exhibit hot downflows, which may fit in the observational scenario of quiet Sun regions \citep{Chae_1998ApJS..114..151C, Brosius_2007ApJ...656L..41B}, further supporting the present conclusions. In \citet{Antolin_2009arXiv0910.0962v1} we have found that Alfv\'en wave heating is effective only in thick loops (with area expansions between photosphere and corona higher than 600) and long loops (with lengths of the corona above 50 Mm), a scenario which may not fit in active regions, where loops exhibit low area expansions due to the high magnetic field filling factors in those regions. Hence Alfv\'en waves may play an important role in the heating of Quiet Sun regions (rather than active regions), where loops are often long, expand more than in active regions, kG (or higher) bright points are ubiquitous and coronal rain appears to be absent.

\acknowledgments

P. A. would like to thank M. Carlsson, V. Hansteen, T. J. Okamoto, L. Heggland, N. Nishizuka, R. Erd\'elyi, K. Ichimoto, H. Isobe, T. Yokoyama and S. Tsuneta for many fruitful discussions. P. A. would also like to acknowledge S. F. Chen for patient encouragement. This work was supported by the Grant-in-Aid for the Global COE Program ``The Next Generation of Physics, Span from Universality and Emergence'' from the Ministry of Education, Culture, Sports, Science and Technology ( MEXT ) of Japan and by a Grant-in-Aid for Creative Scientific Research, ``The Basic Study of Space Weather Prediction'' (17GS0208; Head Investigator: K. Shibata), from the Ministry of Education, Science, Sports, Technology, and Culture of Japan. The numerical calculations were carried out on Altix3700 BX2 at YITP in Kyoto University.

\bibliographystyle{aa}
\bibliography{aamnemonic,patbib}  

\clearpage

\input{table1}

\end{document}